\documentclass[twocolumn]{aastex62}

\usepackage{array}
\usepackage{amsmath}
\usepackage[counterclockwise, figuresleft]{rotating}

\begin{document}


\title{Exoplanet Imitators: A test of stellar activity behavior in radial velocity signals}

\author[0000-0001-8838-3883]{Chantanelle Nava}
\affiliation{Center for Astrophysics ${\rm \mid}$ Harvard {\rm \&} Smithsonian, 60 Garden Street, Cambridge, MA 02138, USA}

\author[0000-0003-3204-8183]{Mercedes L\'{o}pez-Morales}
\affiliation{Center for Astrophysics ${\rm \mid}$ Harvard {\rm \&} Smithsonian, 60 Garden Street, Cambridge, MA 02138, USA}

\author[0000-0001-9140-3574]{Rapha\"{e}lle D. Haywood}
\affiliation{Center for Astrophysics ${\rm \mid}$ Harvard {\rm \&} Smithsonian, 60 Garden Street, Cambridge, MA 02138, USA}
\affiliation{NASA Sagan Fellow}

\author[0000-0001-6777-4797]{Helen A.C. Giles}
\affiliation{Observatoire de Gen\`{e}ve, Universit\'{e} de Gen\`{e}ve, Chemin des Maillettes 51,1290 Versoix, Switzerland}




\begin{abstract}
Accurately modeling effects from stellar activity is a key step in detecting radial velocity signals of low-mass and long-period exoplanets. Radial velocities from stellar activity are dominated by magnetic active regions that move in and out of sight as the star rotates, producing signals with timescales related to the stellar rotation period. Methods to characterize radial velocity periodograms assume that peaks from magnetic active regions will typically occur at the stellar rotation period or a related harmonic. However, with surface features unevenly spaced and evolving over time, signals from magnetic activity are not perfectly periodic, and the effectiveness of characterizing them with sine curves is unconfirmed. With a series of simulations, we perform the first test of common assumptions about signals from magnetic active regions in radial velocity periodograms. We simulate radial velocities with quasi-periodic signals that account for evolution and migration of magnetic surface features. As test cases, we apply our analysis to two exoplanet hosts, Kepler-20 and K2-131. Simulating observing schedules and uncertainties of real radial velocity surveys, we find that magnetic active regions commonly produce maximum periodogram peaks at spurious periods unrelated to the stellar rotation period: 81\% and 72\% of peaks, respectively for K2-131 and Kepler-20. These unexpected peaks can potentially lead to inaccuracies in derived planet masses. We also find that these spurious peaks can sometimes survive multiple seasons of observation, imitating signals typically attributed to exoplanet companions. \\
\end{abstract}

\section{Introduction}\label{Introduction}

Current radial velocity (RV) observations aim to detect signals of less massive and/or longer period planets than ever before. These planets induce RV semi-amplitudes similar to or smaller than those produced by stellar activity \citep[e.g.][]{LMH, K2-131, haywood18}. The field has reached an era, therefore, in which understanding and correcting for stellar activity is essential to accurate RV detections and characterizations of interesting new exoplanets. \\ 

 Stellar signals in RVs result from three main physical processes: variations in a star's internal pressure produce surface oscillations; convection leads to surface granulation; and magnetic activity produces surface spots, plage, and faculae \citep{Fischer16}. Oscillation effects are dominated by p-modes that occur on timescales of minutes, while granulation effects occur on timescales of minutes to hours \citep{leighton62, labonte81, kuhn83}.  When observed over three 10-minute exposures in a night, each separated by approximately two hours, signals from p-mode oscillations and granulation can average to less than one meter per second (m/s) for Sun-like stars \citep{dumusque11, meunier15, Chaplin19}. These observing strategies alleviate many, but not all effects of stellar activity on RVs, with at least meter-per-second signals from magnetic active regions remaining \citep[e.g.][]{makarov2009, meunier2010, lagrange2011, haywood14}. As the next generation of spectrographs come online, precision of RV mass determinations will not be limited by astronomical instruments, but rather by our ability to model and remove signals from magnetic active regions.\\
 
 Magnetic active regions containing spots, plage, and faculae impact the overall flux measured from their location. As the star rotates, active regions on the approaching limb impact the amount of blue-shifted light measured, and those on the receding limb impact the amount of red-shifted light measured. Magnetic active regions also suppress convective blue-shift at any location to produce a net red-shifted effect \citep{meunier10a, meunier10b, Haywood16}. RV signals from magnetic active regions vary quasi-periodically as activity features evolve on the rotating stellar surface. These signals can make detecting and characterizing exoplanets particularly difficult when the planets' orbital periods fall close to the rotation period of the star (\emph{$P_{\rm{rot}}$}). \cite{newton16_prot} and \cite{vanderburg16} demonstrated this to be a potentially serious challenge in the case of exoplanets orbiting in the habitable zones of M-dwarf stars.\\
 
 The interference of magnetic activity signals in our ability to detect exoplanet RV signals is a long known problem \citep[e.g.][]{Maxted2011, Gillon2011}. It is customary, therefore, to estimate $P_{\rm{rot}}$ using a variety of methods.
\emph{$P_{\rm{rot}}$} can be estimated using the equation, $P_{\rm{rot}} \approx 2\pi R_{*} / vsini$, where \emph{$R_{*}$} is the radius of the star, \emph{v} is its rotational velocity, and \emph{i} is the inclination of the star's rotation axis with respect to Earth. \emph{$R_{*}$} can be accurately measured using asteroseismology \citep{Bedding2010, Huber2013}, interferometry \citep{Boyajian12a, Boyajian12b}, or stellar spectral models. Empirical relations between stellar effective temperature and radius have also been established to estimate radii of main sequence A, G, F, and M-dwarf stars \citep{boyajian12, mann15}. The value of $vsini$ can be measured from the width of spectral lines. However, \emph{i} is typically unknown, and therefore the value of \emph{$P_{\rm{rot}}$} derived using this method is only an upper limit. Another method to estimate \emph{$P_{\rm{rot}}$} utilizes the empirical relation, derived by \cite{wright11}, between a star's X-ray bolometric luminosity and $P_{\rm{rot}}$. However, this method depends on many simplifying assumptions about stellar dynamos and is susceptible to systematic errors associated with pre-main sequence and binary stars. These assumptions and errors lead to non-quantifiable uncertainties in final \emph{$P_{\rm{rot}}$} estimates. \\

\emph{$P_{\rm{rot}}$} can also be derived from a star's photometric light curve (LC). The performance of this method has greatly improved as data from dedicated, high-cadence space-based and ground-based photometric surveys have become available. For example, \cite{mcquillan14} estimated \emph{$P_{\rm{rot}}$} for main-sequence stars using Kepler mission LCs, \cite{haywood14} demonstrated one of the early applications of a Gaussian Processes (GP) regression model to constrain \emph{$P_{\rm{rot}}$} from a Kepler LC, and \cite{newton16_prot} estimated \emph{$P_{\rm{rot}}$} for 387 nearby M-dwarfs using MEarth LCs \citep{irwin2009}. \\


The method used by \cite{mcquillan14} utilizes auto-correlation functions (ACF) to estimate \emph{$P_{\rm{rot}}$}. However, the error associated with this method is based solely on chi-squared fits and does not account for simplified assumptions about the evolution and distribution of magnetic active regions. Additionally, many LCs show no clear rotational signals and produce inconclusive ACF results. GP regression models, like those used by \cite{haywood14}, avoid deterministic functions and can generate complex signals from magnetic activity observed in RVs. GP regression is currently the most physically motivated method to model stellar activity. \cite{Angus18} demonstrated that GP regression with a quasi-periodic kernel provides more accurate estimates of \emph{$P_{\rm{rot}}$} from LCs than both sine-fitting with periodograms and auto-correlation function analyses. However, it is computationally intensive and rarely applied to high-cadence data sets. \cite{newton16_prot} estimated \emph{$P_{\rm{rot}}$} according to the statistically significant periodogram peak resulting in the best-fit sine curve to the transit-removed LC. The error associated with this method suffers from the same issues as that of ACF analysis, and many M-dwarfs produce incoherent LCs that cannot
be characterized by a simple sine curve.\\

RV analyses rely heavily on \emph{$P_{\rm{rot}}$} estimates from the methods discussed above to differentiate magnetic activity signals from Keplerian signals produced by exoplanets. As long as a statistically significant peak in the RV periodogram is located more than a few days from the estimated value of \emph{$P_{\rm{rot}}$}, it will often be explored as a potential exoplanet signal. If the peak is \emph{long-lived}, surviving over multiple seasons of observation, stellar activity is considered unlikely to be the source of the signal \citep[e.g.][]{Kep20, Pinamonti19}. \\

Validation of exoplanet companions with the above method relies on a number of assumptions. First, it assumes that estimates of \emph{$P_{\rm{rot}}$} from other methods are correct within a few days. Next, it assumes that magnetic activity signals will usually produce a maximum peak in the periodogram near \emph{$P_{\rm{rot}}$} or one of its major harmonics. Finally, it assumes that signals from evolving activity features will not produce \emph{long-lived} peaks in the RV periodogram at periods unrelated to \emph{$P_{\rm{rot}}$}. \\

In this paper we test the last two assumptions above by analyzing how magnetic activity signals present in periodograms of real RV data. We describe our methods in Section \ref{methods}, and in Section \ref{analysis}, as test cases, we apply them to the known planetary systems, Kepler-20 and K2-131. In Section \ref{results}, we report our results and in Section \ref{discussion}, we discuss the implications of those results with respect to reliable RV detection of exoplanets in the presence of stellar activity signals.

\newpage
\section{Method}\label{methods}

 Our method follows three main steps: simulate magnetic activity RV signals, select model parameters, and investigate simulated RV periodograms. Here we provide a general outline of our method. \\

\subsection{Simulation of Magnetic Activity RV Signals}\label{GPs}

We simulate magnetic activity RV signals using GP regression with a quasi-periodic kernel, motivated by the work of \cite{haywood14}. The quasi-periodic kernel has the form:

\begin{equation}
\begin{split}
 k(t, t') = A^2  \exp\bigg[-\frac{(t-t')^2}{2\tau^2}\\
 - \frac{2sin^2\left(\frac{\pi (t-t')}{P_{\rm{rot}}}\right)}{\omega^2}\bigg],
 \end{split}
\end{equation} \label{GP_eq}

where \emph{k(t,t')} is the correlation weight between observations taken at times t and t'. \emph{A} is the mean amplitude of the activity signal, \emph{$\tau$} is related to the evolution timescale of activity features, \emph{$P_{\rm{rot}}$} is the stellar rotation period, and \emph{$\omega$} is related to the average distribution of activity features on the surface of the star. The parameter \emph{$\omega$} describes the level of high-frequency variation expected within a single stellar rotation. Since the level of high-frequency variation defines the number and spacing of local minima or maxima within the timescale of \emph{$P_{\rm{rot}}$}, it is physically related to the average distribution of magnetic active regions on the stellar surface. For each unique set of GP hyper-parameters (\emph{A}, \emph{$\tau$}, \emph{$P_{\rm{rot}}$}, \emph{$\omega$}), we simulate 100,000 iterations of stellar RV signals by sampling randomly from the GP prior distribution, with each iteration representing a different phase of the activity signal.\\

Using Equation (1), we model magnetic activity RVs for targets observed by current RV campaigns, assigning observation times and uncertainties from the real RV data with which we later compare our modeled RVs. We apply a bootstrapping method to real RV uncertainties, using them in a different randomized order with each new iteration of modeled RVs. Section \ref{analysis} details observation times and RV uncertainties for specific test targets.\\

\subsection{Selection of Model Parameters}\label{params}

The values of \emph{A}, \emph{$\tau$}, and \emph{$P_{\rm{rot}}$} for a given target are adopted from the literature when available, or are estimated as follows. We set \emph{A} equal to the standard deviation of the target's real RV residuals (with signals from confirmed exoplanets removed), minus the median value of reported observational uncertainties. With this value, we test a case in which any remaining spread in the RVs, after the removal of known exoplanet signals, can be attributed to a combination of stellar activity and errors associated with observations. \cite{rajpaul16} showed how the removal of signals from known exoplanets can lead to spurious periodic signals in RV data sets. However, only the mean values of our modeled RV data sets (\emph{A}) depend on real RV residuals. The overall structure of each of our modeled data sets are independent of the value of \emph{A}, and therefore are insusceptible to the spurious periodic signals mentioned above. We set \emph{$\tau$} and \emph{$P_{\rm{rot}}$} according to best estimates from LC and/or RV analyses performed on the data sets. Section \ref{analysis} details the values of \emph{A}, \emph{$\tau$}, and \emph{$P_{\rm{rot}}$} adopted for specific targets. \\

For the purpose of our tests, we use two values of \emph{$\tau$} to probe different relationships between stellar rotation and the evolution timescale of magnetic active regions. In the first, with \emph{$\tau$} $\approx$ \emph{$P_{\rm{rot}}$}, we utilize real \emph{$\tau$} estimates from LC and/or RV analyses mentioned above. In the second, with \emph{$\tau$} = 10 \emph{$P_{\rm{rot}}$}, we explore the case of highly \emph{stable} magnetic activity features, compared to measured activity lifetimes on Sun-like stars \citep{giles17}. This is the case of an unchanging stellar surface over multiple rotations and timescales of typical RV observations. Some faculae regions fall under this category, with features surviving up to ten times as long as spots \citep{ACC19}.    \\

To explore whether uncertainties associated with hyper-parameter estimates affect final maximum peak distributions, we performed additional simulations using A, \emph{$\tau$} and \emph{$P_{\rm{rot}}$} values falling at the high and low limits of their computed uncertainties. In most of these test cases, final distributions and occurrence rates had similar overall trends to simulations using our published hyper-parameter values. Any exceptions to this are further discussed in Section \ref{analysis}.\\

As mentioned above, \emph{$\omega$} is physically related to the average distribution of magnetic active regions on the stellar surface. Models have demonstrated that even highly complex activity distributions will average to just two to three large active regions in a given rotation \citep{jeffers}. The distribution \emph{$\omega$} $= 0.5 \pm 0.05$ is consistent with this behavior, allowing for two to three local minima or maxima per rotation. This prior on \emph{$\omega$} has been used successfully to determine exoplanet masses from a number of RV data sets \citep[e.g.][]{haywood14, grunblatt15, LMH, haywood18}. While several RV characterizations have used broader priors on \emph{$\omega$}, the results from \cite{jeffers} make a strong case for the much tighter Gaussian prior above \citep[e.g.][]{mortier16, faria16, cloutier17, AD17}. Broader priors risk over-fitting other noise signals, and mistakenly attributing them as part of the magnetic activity signal. We simulate five different values of \emph{$\omega$} for each target, sampling evenly from the above distribution, i.e. \emph{$\omega$} = [0.45, 0.475, 0.5, 0.525, 0.55].  \\

\subsection{Investigation of Simulated RV Periodograms}\label{maxpeak} 

In each simulation, we investigate distributions of maximum RV periodogram peaks locations over 100,000 iterations. In each iteration, we generate an RV signal and calculate a Generalized Lomb-Scargle (GLS) periodogram on the signal \citep{Lomb, Scargle, GLS}. We calculate the GLS periodogram with a lower limit of 1.5 days (to avoid the 1-day peak due to nightly observations) and an upper limit of half the baseline of the observations' time span (to consider only periods detectable in the simulated data). From each GLS periodogram, we record the period of the maximum statistically significant peak, with a false alarm probability (FAP) rate $>1\%$. We repeat the process of generating modeled RVs and identifying maximum periodogram peaks over a number of iterations, plotting the final distribution in a histogram. \\

We consult the final distribution of peak periods to calculate how often maximum peaks in the periodogram occur at a series of important periods, detailed below. We define two \emph{occurrence rates} at any given period: the first is the percentage of iterations with significant (FAP $> 1\%$) peaks in the RV periodogram that have a maximum peak falling within 5$\%$ of that period, and the second is the percentage of all 100,000 iterations that have a maximum peak falling within 5$\%$ of the period. We base the 5$\%$ metric on the range of uncertainties produced by LC estimates of \emph{$P_{\rm{rot}}$}, used as priors in RV fits \citep[e.g.][]{LMH, Kep20, K2-131}. \\

We calculate occurrence rates at periods related to \emph{$P_{\rm{rot}}$}, including \emph{$P_{\rm{rot}}$} itself and integer multiples of \emph{$P_{\rm{rot}}$} up to the longest period for which the periodograms were calculated. These periods also include rotational harmonics (e.g. \emph{$P_{\rm{rot}}$}/2, \emph{$P_{\rm{rot}}$}/3). We calculate occurrence rates for the same number of rotational harmonics as calculated for integer multiples of \emph{$P_{\rm{rot}}$}. For example, if we calculate occurrence rates for integer multiples up to $P_{\rm{rot}}\times$5, we calculate occurrence rates for rotational harmonics down to \emph{$P_{\rm{rot}}$}/5. To track the window function signal, we also calculate occurrence rates for the period of the \emph{cadence peak}, the maximum peak produced by the cadence of observations. To calculate the cadence periodogram, we produce a signal with the same time stamps as the observations and replace RV amplitudes by random values from the uniform distribution $1.0 \pm 1\times10^{-15}$. We then calculate a GLS periodogram on the cadence signal with the same period limits used to calculate the simulated RV periodograms. \\

Finally, we calculate occurrence rates for maximum peaks falling at a specified period of interest (POI), typically the period of an exoplanet candidate in question. Section \ref{analysis} details POI selections for specific targets. For targets with multiple seasons of simulated RVs, we also determine a rate of \emph{time-coherence} at the POI. In iterations with a maximum peak occurring at the POI, we calculate the GLS periodogram of each independent season by setting unused RVs to zero with an error of 100 m/s. This method preserves periodic signals inherent to the observational cadence and was first described in \cite{Dumusque12}. We define a maximum periodogram peak at the POI to be \emph{long-lived} if it remains the maximum peak in each of the periodograms of each individual season. We define the rate of \emph{time-coherence} as the number of iterations with a long-lived maximum peak occurring at the POI divided by the total number of iterations with a maximum peak at the POI. \\



\section{Application to K2-131 and Kepler-20}\label{analysis}

We used the method described above to explore potential effects of magnetic activity in the published RV measurements of two known exoplanet systems: K2-131 and Kepler-20. Both systems contain planets detected via the transit method, with additional strong periodic signals detected in follow-up RVs and considered as potential non-transiting exoplanet companions.

\subsection{K2-131}\label{K2_analysis}

\begin{figure*}[]
\epsscale{1.0}
\plotone{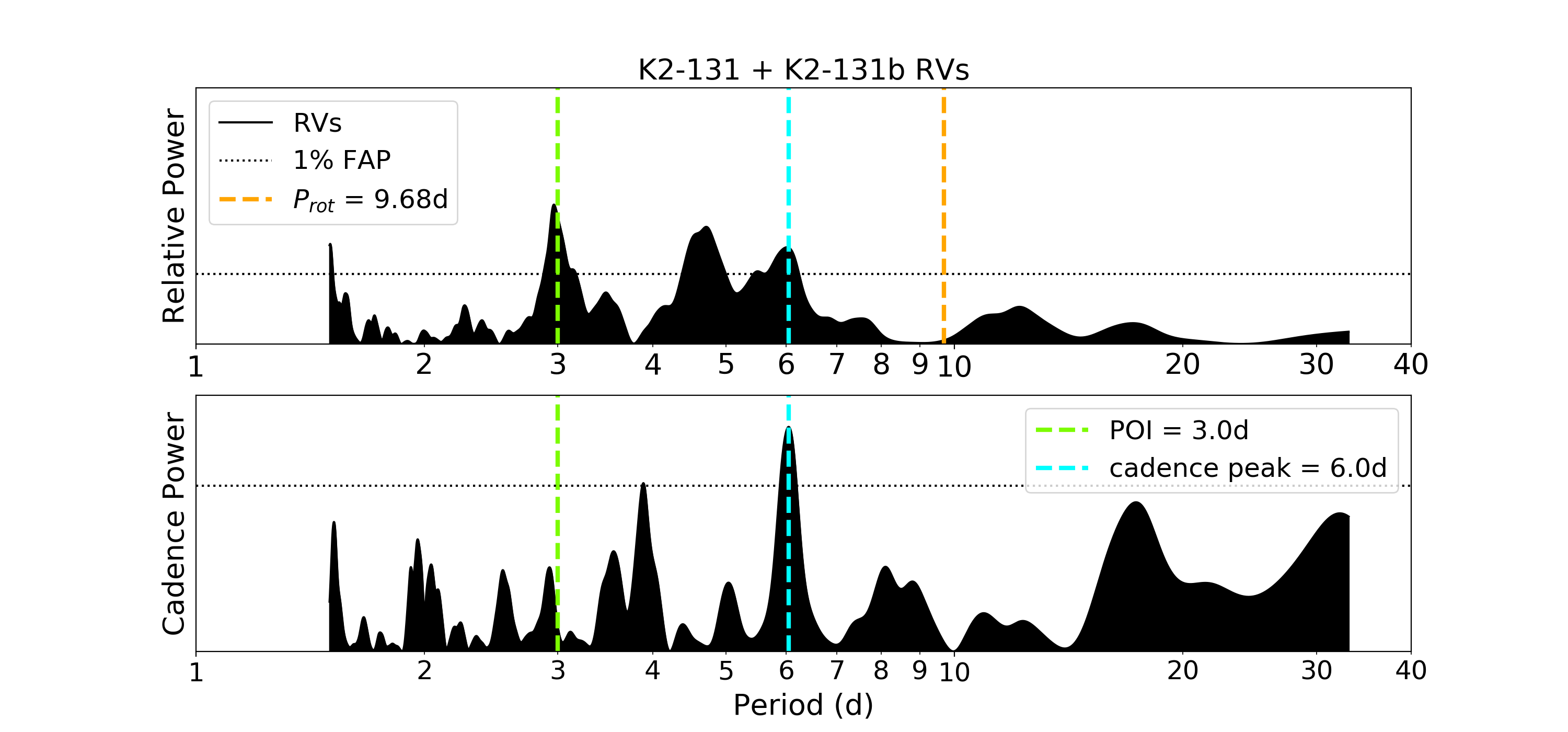}
\caption{Top: The GLS periodogram of K2-131's combined HARPS-N/PFS RVs. \cite{K2-131} explored the strong signal at the period of interest (\emph{POI}), 3.0 days, as potential evidence of a non-transiting exoplanet companion. Bottom: The periodogram inherent to the cadence of K2-131 RV observations. The maximum peak inherent to observational cadence, the \emph{cadence peak}, exists at 6.0 days.}\label{K2_splt}
\end{figure*}

\begin{figure*}[]
\includegraphics[scale=0.45]{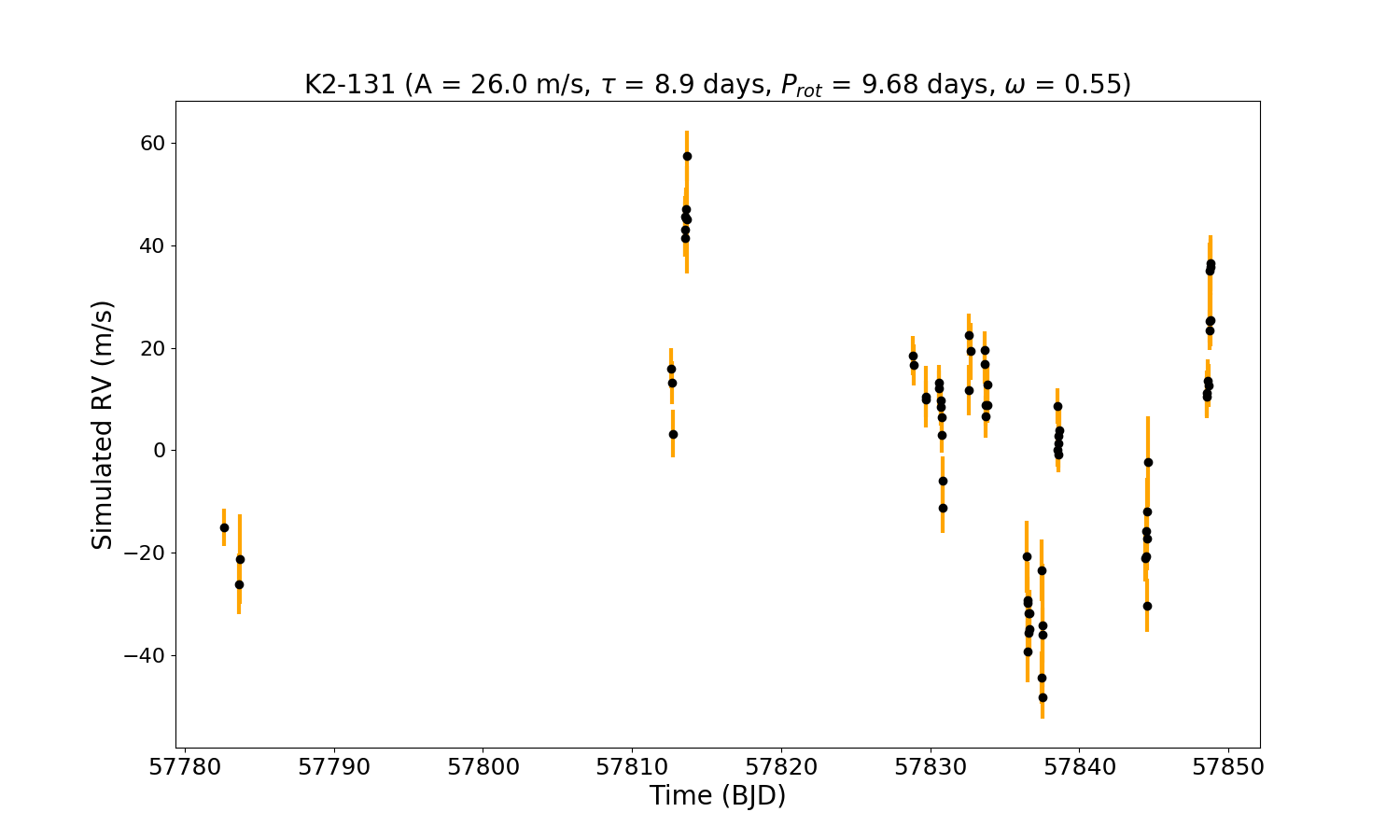}
\caption{An example simulated RV data set for the combined magnetic activity signal of K2-131 and its companion K2-131b, using observation times from combined HARPS-N/PFS RVs \citep{K2-131}.}\label{K2-131_scat}
\end{figure*}

K2-131 is a solar-type star with \emph{$P_{\rm{rot}}$} = 9.68 days and one ultra-short period transiting exoplanet companion, K2-131b ($P_{b}$ = 0.369 days), confirmed by \cite{K2-131} with combined RV observations from HARPS-N \citep{Cosentino2012} and the Magellan Planet Finder Spectrograph \citep[PFS,][]{Crane2010}. In addition to the RV signal of K2-131b, \cite{K2-131} found a statistically significant peak at 3.0 days in the periodogram of their combined PFS/HARPS-N RV observations, reproduced here in Figure \ref{K2_splt}. The peak raised the possibility of a potential additional non-transiting exoplanet in the system. However, they attributed the signal to magnetic activity due to its close proximity to the second rotational harmonic (\emph{$P_{\rm{rot}}$}/3 = 3.2 days) and additional detection of the signal in data from two magnetic activity indicators.  \\

We simulated combined RV signals of K2-131 and its known transiting exoplanet, K2-131b. We set a value of POI = 3.0 days to investigate whether the combined signal from K2-131's magnetic activity and K2-131b could produce the 3.0-day periodic signal discussed above. We used observation times and RV uncertainties from the combined set of 41 HARPS-N and 32 PFS observations of K2-131, in which the 3.0-day signal was originally detected \citep[Figure \ref{K2-131_scat},][]{K2-131}. All the observations were taken within a single season, the HARPS-N data between January and April 2017, and the PFS data over six nights in March and April 2017. The observations have a baseline of approximately 66 days, so we set the upper limit for detection of a periodic signal at one-half that time span, approximately 33 days.  \\

We simulated magnetic activity RVs of K2-131 using GP regression and the quasi-periodic kernel described in Equation (1). In their original RV analysis, \cite{K2-131} utilized GP regression to fit for magnetic activity. We used the final GP hyper-parameters reported in Table 7 of their paper: A = 26.0 m/s, \emph{$P_{\rm{rot}}$} = 9.68 days, \emph{$\tau$} = 8.9 days. We test this set of hyper-parameters with each of the five average activity distributions, \emph{$\omega$} = [0.45, 0.475, 0.50, 0.525, 0.55]. We also ran a second set of simulations with the activity evolution timescale increased to \emph{$\tau$} = 96.8 days, in order to explore the case of activity features that remain stable over the timescale of observations. \\

As described in Section \ref{params}, we performed additional simulations using A, \emph{$\tau$} and \emph{$P_{\rm{rot}}$} values falling at the high and low limits of the uncertainties reported in Table 7 of \cite{K2-131}. Most of these test cases yielded final occurrence rates and overall trends similar to those reported in Section \ref{results}. However, cases testing the low limit of \emph{$P_{\rm{rot}}$} have an overlap in values at the POI and \emph{$P_{\rm{rot}}$}/3 (3.0 days and 3.18 days, respectively), within the 5\% error. In these cases, occurrence rates at the POI increased to resemble values reported for \emph{$P_{\rm{rot}}$}/3 in Tables \ref{K2-131_PPev_rates} and \ref{K2-131_Lev_rates}.  \\

In all cases, we simulated the planetary RVs of K2-131b with a Keplerian signal, assigning the exoplanet parameters reported in table 7 of \cite{K2-131}: $K = 6.55$ m/s, $P = 0.369$ days, $e = 0$, $t_{c} = 3582.9360$ (BJD - 2,454,000), where $K$ is the semi-amplitude of the exoplanet RV signal, $P$ is the orbital period, $e$ is the orbital eccentricity, and $t_{c}$ is the central time of transit. With our lower period limit for periodogram calculations set to 1.5 days, we did not track maximum peak occurrences at the orbital period of K2-131b ($P = 0.369$ days) in subsequent analysis of simulated magnetic activity signals. However, since the signal at 3.0 days appeared in real RV data for K2-131 before the removal of the signal from K2-131b, we included the planetary signal to investigate the possibility of the combined signals from magnetic activity and K2-131b producing a maximum peak at the POI.  \\

\begin{figure*}[!]
\includegraphics[scale=0.5]{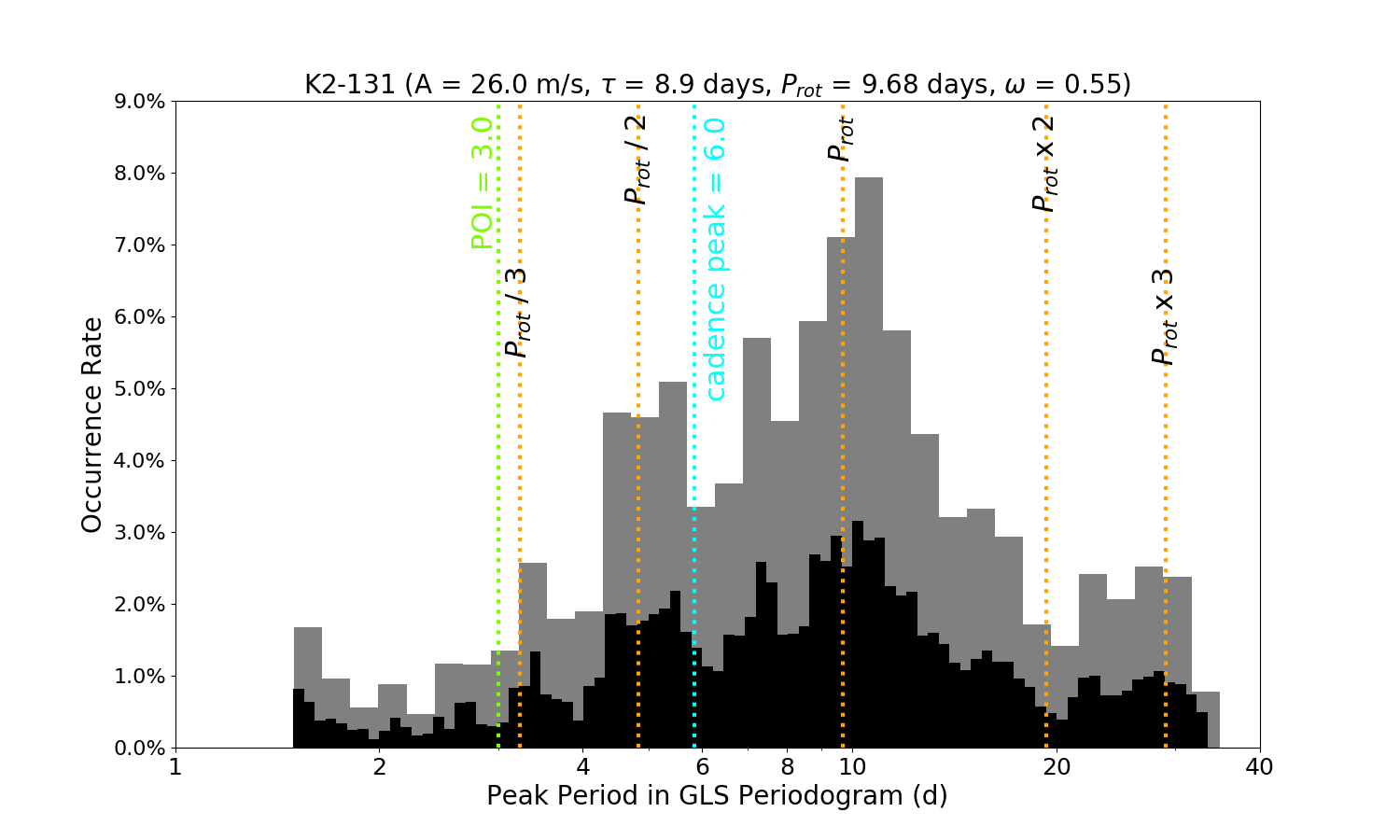}
\caption{A histogram of maximum peak periods for simulated combined RVs of K2-131 and K2-131b, with evolving activity features (\emph{$\tau$} = 8.9 days) and \emph{$\omega$} = 0.55. The histogram is shown in grey with the bin size set to match occurrence rates listed in Table \ref{K2-131_PPev_rates}, and shown with a smaller bin size in black to show more detail. The vertical lines correspond to periods for which occurrence rates were calculated.}\label{K2-131_PPev_dist}
\end{figure*}

\begin{figure*}[!]
\includegraphics[scale=0.5]{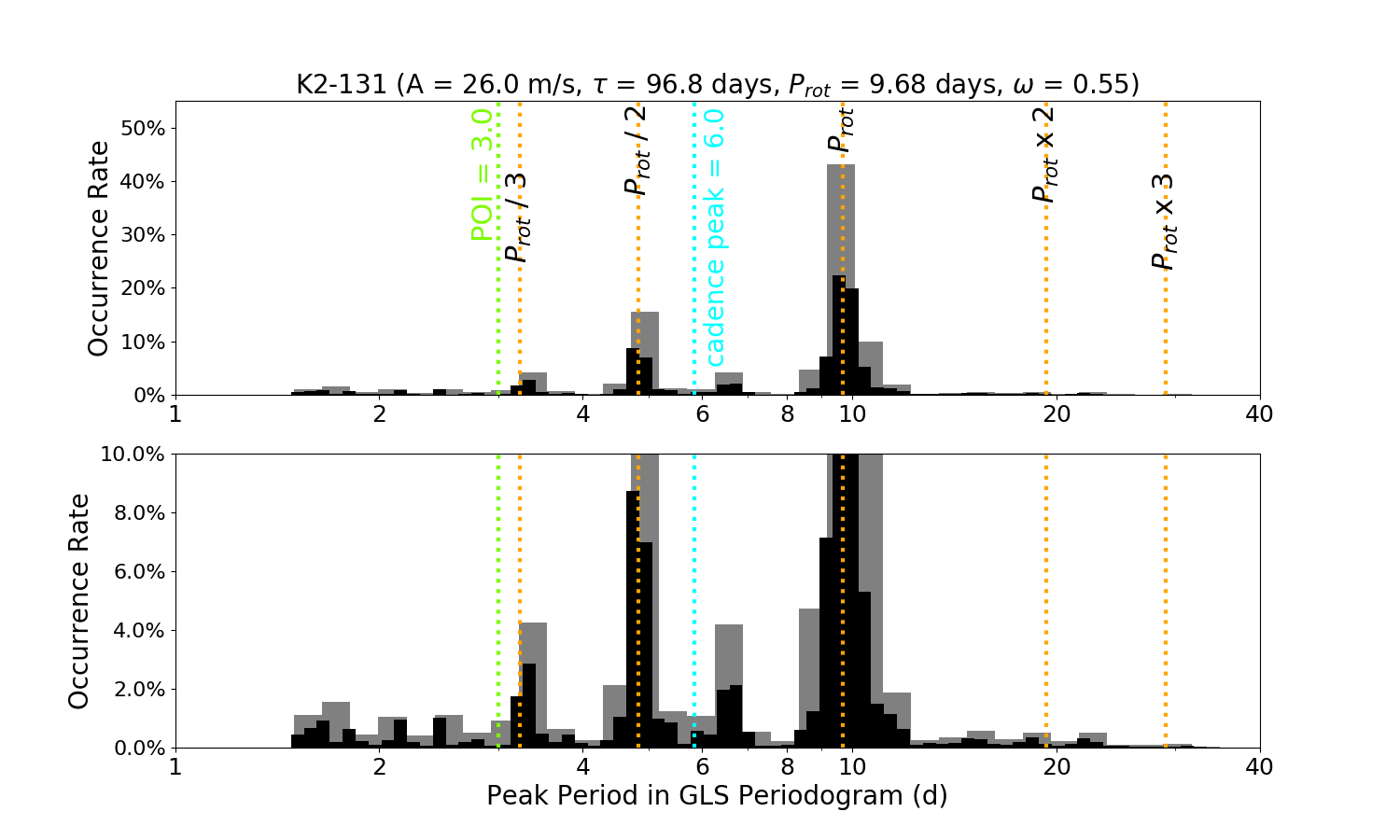}
\caption{The same as shown in Figure \ref{K2-131_PPev_dist}, but with stable activity features (\emph{$\tau$} = 96.8 days) and occurrence rates in grey from Table \ref{K2-131_Lev_rates}. The top panel shows the full distribution and the bottom panel shows a zoomed in view.}\label{K2-131_Lev_dist}
\end{figure*}

We generated a total of ten simulations, using the two values of \emph{$\tau$} and five values of \emph{$\omega$} given above, with \emph{A} and \emph{$P_{\rm{rot}}$} fixed. In each simulation, we generated 100,000 iterations of RVs from K2-131 and its known exoplanet companion. Figures \ref{K2-131_PPev_dist} and \ref{K2-131_Lev_dist} show example histograms for simulations with \emph{$\tau$} = 8.9 days and \emph{$\tau$} = 96.8 days, respectively, both with \emph{$\omega$} = 0.55. All five values of \emph{$\omega$} produced similar final maximum peak distributions, but for simplicity, we only show distributions for a single value of \emph{$\omega$}. Results for the other values of \emph{$\omega$} are reported in Tables \ref{K2-131_PPev_rates} and \ref{K2-131_Lev_rates}.\\

Tables \ref{K2-131_PPev_rates} and \ref{K2-131_Lev_rates} list occurrence rates for simulations with \emph{$\tau$} = 8.9 days and \emph{$\tau$} = 96.8 days, respectively. We calculated occurrence rates for maximum peaks located within 5$\%$ of \emph{$P_{\rm{rot}}$}, its rotational harmonics (\emph{$P_{\rm{rot}}$}/2, \emph{$P_{\rm{rot}}$}/3), its integer multiples ($P_{\rm{rot}}\times$2, $P_{\rm{rot}}\times$3), the cadence peak (6.0 days, Figure \ref{K2_splt}), and the POI (3.0 days, Figure \ref{K2_splt}). With only a single season of RVs available in the original data set, we did not calculate a rate of time-coherence at the POI. \\


\begin{deluxetable*}{cccccc}
\tabletypesize{\scriptsize}
\tablewidth{0pt}
\tablecolumns{7}
\tablecaption{Occurrence rates (\%) for select maximum peak values in periodograms of simulated combined RVs from K2-131 and K2-131b, with evolving activity features (\emph{$\tau$} = 8.9 days). Only 13 iterations of simulated RV periodograms had no statistically significant peaks, so occurrence rates with respect to just iterations having statistically significant peaks and with respect to all 100,000 iterations were equal in the case of K2-131. Therefore, only a single occurrence rate is listed for each combination of \emph{$\omega$} and period values. \label{K2-131_PPev_rates}}
\tablehead{\colhead{Peak value}  & \colhead{\emph{$\omega$} = 0.45} & \colhead{\emph{$\omega$} = 0.475}  & \colhead{\emph{$\omega$} = 0.5}  & \colhead{\emph{$\omega$} = 0.525}  &  \colhead{\emph{$\omega$} = 0.55}}

\startdata 
POI = 3.0d  & 1.7  & 1.5  & 1.3  & 1.1 & 1.0  \\
\\
\emph{$P_{\rm{rot}}$} = 9.7d  & 6.6  & 6.8  & 7.4   & 7.6  & 7.7   \\
\\
\emph{$P_{\rm{rot}}$} / 2 = 4.8d   & 5.8  & 5.5   & 5.3   & 4.9   & 4.8   \\
\emph{$P_{\rm{rot}}$} / 3 = 3.2d  & 3.6  & 3.3    & 3.0   & 2.7  & 2.5   \\
\\
 \emph{$P_{\rm{rot}}$} x 2 = 19.4d   & 1.3  & 1.4    & 1.3   & 1.3   & 1.5  \\
\emph{$P_{\rm{rot}}$} x 3 = 29.0d   & 2.1   & 2.3   & 2.4  & 2.5 & 2.7  \\
\\
cadence peak = 6.0d & 3.8  & 3.7  & 3.6   & 3.5 & 3.3  \\
\enddata

\end{deluxetable*}

\begin{deluxetable*}{cccccc}
\tabletypesize{\scriptsize}
\tablewidth{0pt}
\tablecolumns{7}
\tablecaption{Occurrence rates (\%) as reported in Table \ref{K2-131_PPev_rates}, but with stable activity features (\emph{$\tau$} = 96.8 days).
\label{K2-131_Lev_rates}}

\tablehead{\colhead{Peak value}  & \colhead{\emph{$\omega$} = 0.45} & \colhead{\emph{$\omega$} = 0.475}  & \colhead{\emph{$\omega$} = 0.5}  & \colhead{\emph{$\omega$} = 0.525}  &  \colhead{\emph{$\omega$} = 0.55}}

\startdata 
POI = 3.0d & 0.4   & 0.4  & 0.3   & 0.3   & 0.2   \\
\\
\emph{$P_{\rm{rot}}$} = 9.7d  & 36.0   & 38.2   & 40.6    & 43.2   & 45.0   \\
\\
\emph{$P_{\rm{rot}}$} / 2 = 4.8d   & 18.3   & 17.9   & 17.6   & 16.7   & 16.2   \\
\emph{$P_{\rm{rot}}$} / 3 = 3.2d  & 6.9   & 6.3    & 5.4   & 4.8   & 4.1   \\
\\
\emph{$P_{\rm{rot}}$} x 2 = 19.4d   & 0.5   & 0.4  & 0.4   & 0.4   & 0.3   \\
\emph{$P_{\rm{rot}}$} x 3 =  29.0d  & 0.1   & 0.1   & 0.1   & 0.1   & 0.1   \\
\\
cadence peak = 6.0d & 1.2   & 1.2   & 1.2   & 1.1   & 1.1   \\
\enddata
\end{deluxetable*}

\subsection{Kepler-20} \label{K20_analysis}

\begin{deluxetable*}{cccccc}
\tabletypesize{\scriptsize}
\tablewidth{0pt}
\tablecolumns{7}
\tablecaption{Transit and orbital parameters for Kepler-20's five transiting exoplanet companions. The majority of values are taken from results reported in Table 4 of \cite{Kep20}. Values marked by a single asterisk are from the fit reported in Table 2 of \cite{gautier}. Values marked by a double asterisk are from the fit reported in Table 1 of \cite{fressin}. \label{transits}}

\tablehead{\colhead{Parameter}  & \colhead{Kepler-20b} & \colhead{Kepler-20c} & \colhead{Kepler-20d}  & \colhead{Kepler-20e}  & \colhead{Kepler-20f}}

\startdata 
orbital period (days)       & $3.696115^{+.000001}_{-.000001}$ & $10.85409^{+.000003}_{-.000003}$ & $77.6113^{+.0001}_{-.0001}$ & $6.098523^{+.000006}_{-.000014}$ & $19.57758^{+.00009}_{-.00012}$   \\
$T_{c}$ (BJD - 2,454,000)   & $967.5020^{+.0003}_{-.0002}$     & $971.6080^{+.0002}_{-.0002}$     & $997.730^{+.001}_{-.002}$   & $968.932^{+.002}_{-.001}$        & $967.5020^{+.0003}_{-.0002}$     \\
orbital eccentricity        & $0.03^{+0.09}_{-0.03}$           & $0.16^{+0.01}_{-0.09}$           &       $<$ 0.6 *  & $<$ 0.28 **  & $<$ 0.32 ** \\
planet radius ($R_{\Earth}$) & $1.868^{+0.066}_{-0.034}$       & $3.047^{+0.084}_{-0.056}$        & $2.744^{+0.073}_{-0.055}$    & $0.865^{+0.026}_{-0.028}$       & $1.003^{+0.050}_{-0.089}$        \\
planet mass ($M_{\Earth}$)   & $9.7^{+1.41}_{-1.44}$           & $12.75^{+2.17}_{-2.24}$          & $10.07^{+3.97}_{-3.70}$      & ...                             & $10.07^{+3.97}_{-3.70}$          \\
\enddata
\end{deluxetable*}

\begin{figure*}
\includegraphics[scale=0.5]{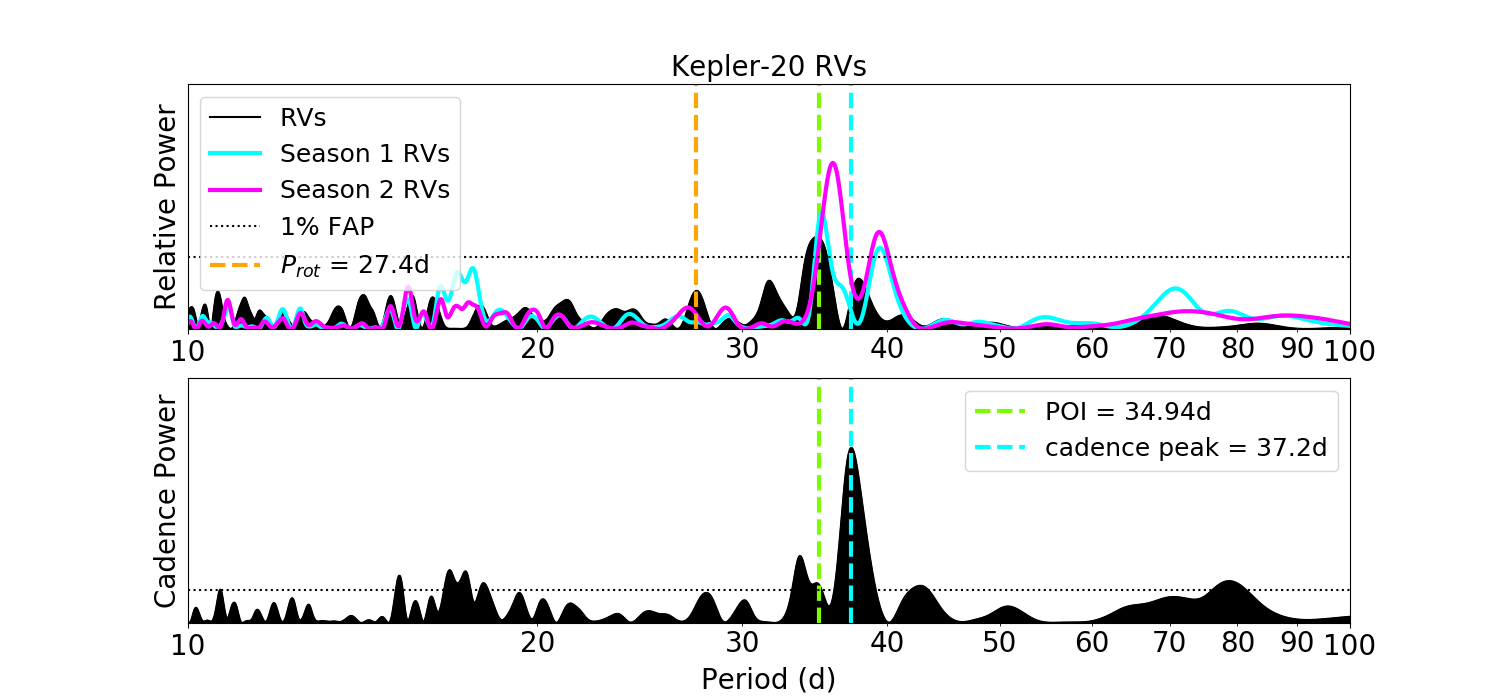}
\caption{Top: The GLS periodogram of Kepler-20's HARPS-N RV residuals (signals from Kepler-20b, Kepler-20c, and Kepler20-d removed) is shown in black. The cyan and green curves show GLS periodograms of just first-season and second-season observations, respectively. The GLS periodogram is calculated for a single season by setting unused RVs to zero with their errors set to 100 m/s, a method first described by \cite{Dumusque12}. The maximum peak at the period of interest (\emph{POI}), 34.94 days, is \emph{long-lived} because it remains the maximum peak in the periodograms of both individual seasons of observations. Due to its long-lived nature, \cite{Kep20} investigated the signal at the POI and attributed it to a non-transiting exoplanet companion. Bottom: The periodogram inherent to the cadence of Kepler-20 RV observations. The \emph{cadence peak} exists at 37.2 days (notably close to the POI at 34.94 days).}\label{K20_splt}
\end{figure*}

\begin{figure*}[]
\includegraphics[scale=0.45]{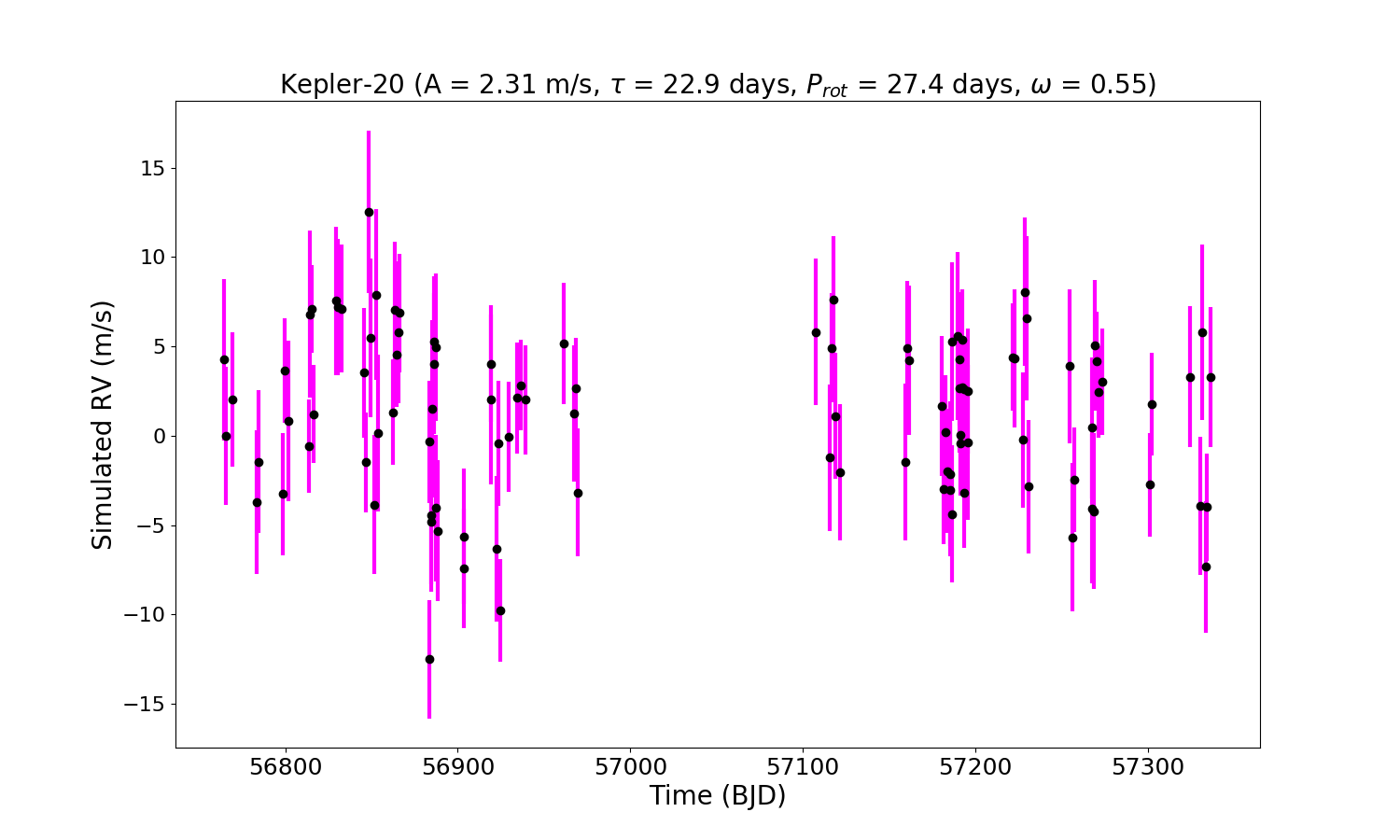}
\caption{An example simulated RV data set for the magnetic activity signal of Kepler-20, using observation times from two seasons of HARPS-N RVs. \citep{Kep20}.}\label{Kep20_scat}
\end{figure*}

Kepler-20 is a solar-type star with \emph{$P_{\rm{rot}}$} = 27.4 days and five confirmed transiting exoplanet companions, all with orbital periods shorter than 80 days \citep[Table \ref{transits},][]{gautier, fressin}. \cite{Kep20} published a sixth non-transiting companion, with combined HARPS-N and HIRES data after removing RV signals from the largest three known transiting planets. The predicted RV amplitudes of signals from the smallest two planets are too small to be detected with HARPS-N and HIRES precision. \cite{Kep20} included the non-transiting planet in their fit after detecting a maximum peak located at 34.94 days in the HARPS-N RV residual periodogram that remained coherent over two seasons of observation separated by approximately 138 days (Figure \ref{K20_splt}). They concluded the long-lived signal to be planetary, considering the approximately 7-day difference between the signal in question and \emph{$P_{\rm{rot}}$} of the star. \cite{Kep20} also observed no correlation between their RVs and the three activity indicators tested, and therefore did not include a model for stellar activity in their final fit. \\

We simulated RVs of Kepler-20's magnetic activity signal. We set a value of POI = 34.94 days to investigate whether the stellar activity signal alone could produce the long-lived periodic signal that \cite{Kep20} attributed to a non-transiting planet, as discussed above. We used observation times and RV uncertainties from the 104 HARPS-N observations of Kepler-20, collected over two seasons between 2014 and 2015 (Figure \ref{Kep20_scat}). We did not include the 30 available HIRES observations because the peak originally motivating the planetary signal at 34.94 days was detected in the periodogram of HARPS-N RVs only. The signal was not detected in the periodogram of combined HARPS-N and HIRES RVs. The HARPS-N observations have a baseline of approximately 570 days, so we set the upper limit for detection of a periodic signal at one-half that time span, approximately 285 days. \\

We simulated magnetic activity RVs of Kepler-20 using GP regression and the quasi-periodic kernel described in Equation (1). Unlike in the case of K2-131, the original RV analysis performed by \cite{Kep20} did not utilize GP regression to fit for magnetic activity, and therefore does not provide estimates of \emph{A}, \emph{$\tau$}, and \emph{$P_{\rm{rot}}$}. We estimated those three hyper-parameters as described below.  \\

We set \emph{A} equal to the standard deviation of the HARPS-N RV residuals minus the median value of the associated uncertainties. We removed signals from Kepler-20b, Kepler-20c, and Kepler-20d to calculate the RV residuals, because it was in these residuals that \cite{Kep20} detected the signal at 34.94 days. As explained before, Kepler-20e and Kepler-20f induce signals smaller in amplitude than HARPS-N RV precision, and therefore can not be reliably fitted and removed \citep{Kep20}. To estimate \emph{$\tau$} and \emph{$P_{\rm{rot}}$}, we performed an ACF analysis on the same Kepler-20 photometric data originally analyzed by \cite{Kep20}, consisting of LCs from fifteen Kepler campaigns (Q3-Q17) collected between 2009 and 2013. We applied discrete shifts to the LC and cross-correlated the shifted LCs with the original, revealing peaks separated by a timescale related to \emph{$P_{\rm{rot}}$}, with correlation powers dropping off at a rate related to \emph{$\tau$} \citep{mcquillan14, giles17}. Our ACF analysis failed to converge on a final value for \emph{$\tau$}, but did produce a $P_{\rm{rot}}$ estimate of $27.4 \pm 0.8$ days. We instead used a value of \emph{$\tau$} = $22.9 \pm 0.2$ days, obtained from the relationship between \emph{$\tau$}, stellar effective temperature, and the scatter in the photometric LC, described in Equation (8) of \cite{giles17}.  We used the following final hyper-parameter values in our reported simulations of Kepler-20: A = 2.31 m/s, \emph{$\tau$} = 22.9 days, \emph{$P_{\rm{rot}}$} = 27.4 days. \\

Again, using the method described in Section \ref{params}, we found that uncertainties associated with our \emph{$\tau$} and \emph{$P_{\rm{rot}}$} estimates did not affect final maximum peak distributions. We also tested two additional values of A: the standard deviation of HARPS-N RV residuals without the mean HARPS-N observational uncertainty subtracted (6.07 m/s) and the reported semi-amplitude of the RV signal attributed to Kepler-20g (4.10 m/s) \citep{Kep20}. In all of these test cases, final distributions and occurrence rates demonstrated similar overall trends to those reported in Section \ref{results}.\\

We ran a second set of simulations with the activity evolution timescale set to \emph{$\tau$} = 274.0 days, in order to explore the case of activity features that remain stable over the timescale of observations. For both sets of A, \emph{$\tau$}, and \emph{$P_{\rm{rot}}$} values, we again tested five average activity distributions, \emph{$\omega$} = [0.45, 0.475, 0.50, 0.525, 0.55].\\

\begin{figure*}[]
\includegraphics[scale=0.5]{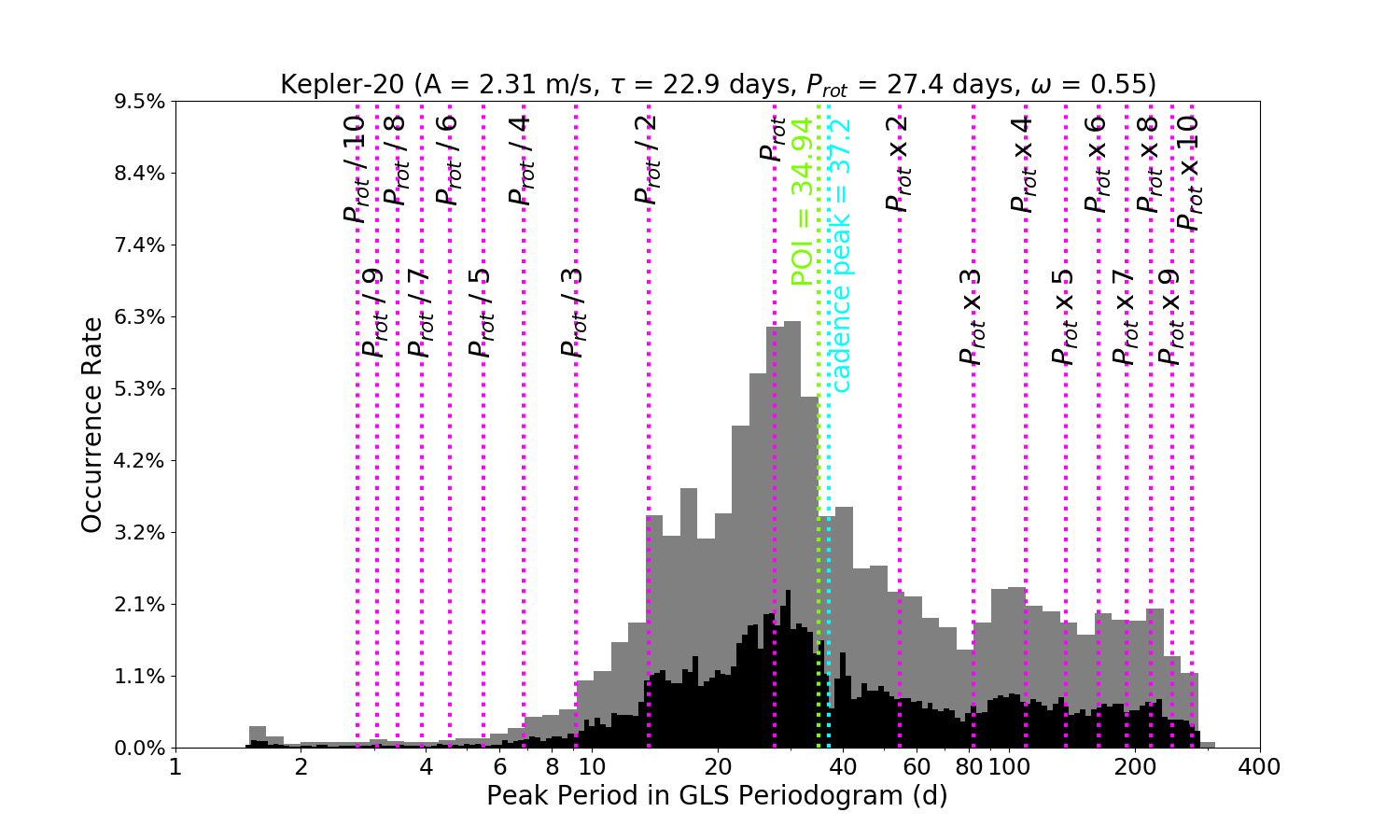}
\caption{A histogram of maximum peak periods for simulated Kepler-20 RVs, with evolving activity features (\emph{$\tau$} = 22.9 days) and \emph{$\omega$} = 0.55. The vertical lines correspond to periods for which occurrence rates were calculated. The histogram is shown in grey with the bin size set to match occurrence rates listed in Table \ref{Kep20_PPev_rates}, and shown with a smaller bin size in black to show more detail. The vertical lines correspond to periods for which occurrence rates were calculated.}\label{Kep20_PPev_dist}
\end{figure*}

\begin{figure*}[]
\includegraphics[scale=0.5]{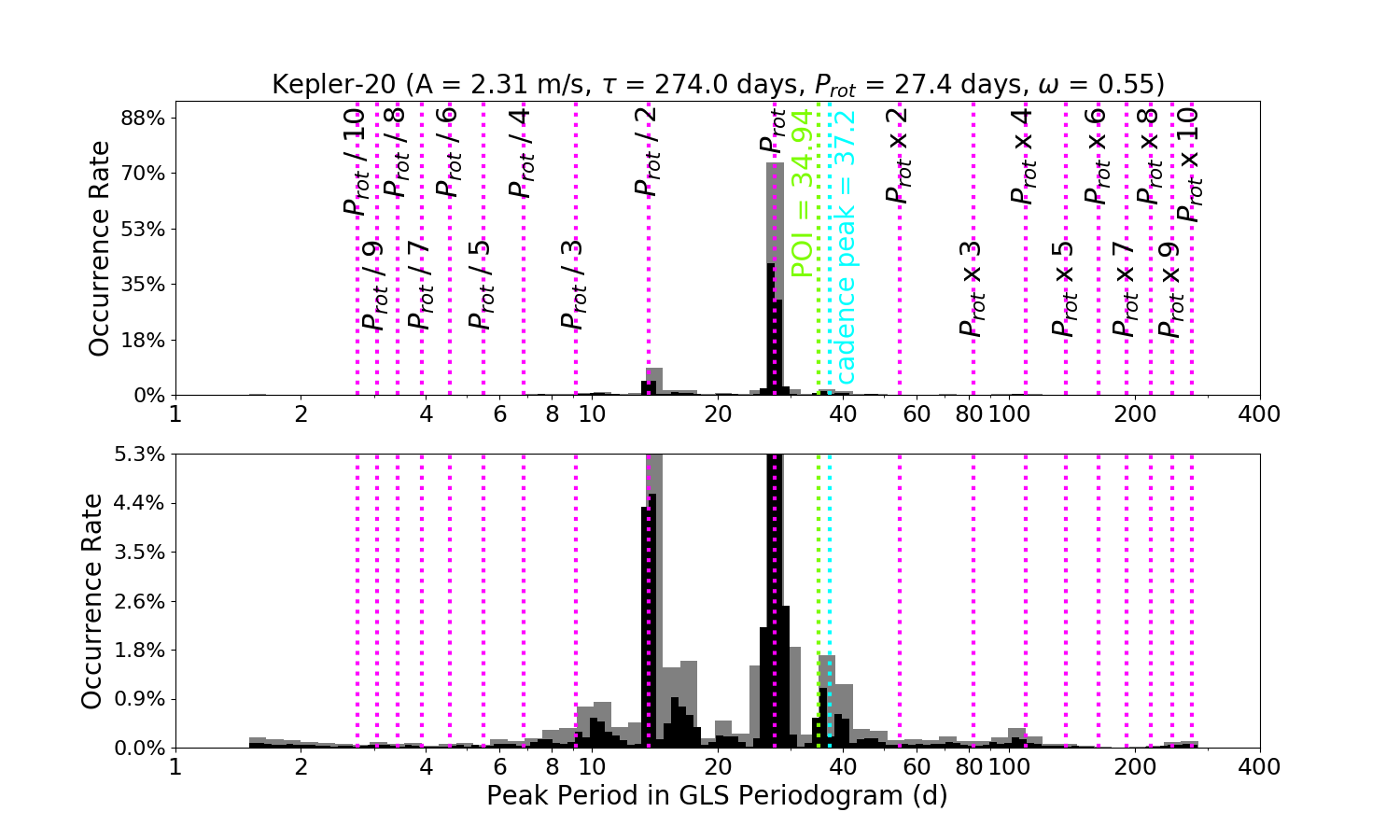}
\caption{The same as shown in Figure \ref{Kep20_PPev_dist}, but with the stable activity features (\emph{$\tau$} = 274.0 days) and occurrence rates in grey from Table \ref{Kep20_Lev_rates}. The top panel shows the full distribution and the bottom panel shows a zoomed in view.}\label{Kep20_Lev_dist}
\end{figure*}

We generated a total of ten simulations, using the two values of \emph{$\tau$} and five values of \emph{$\omega$} given above, again with \emph{A} and \emph{$P_{\rm{rot}}$} fixed. In each simulation, we generated 100,000 iterations of magnetic activity RVs for Kepler-20. We plotted a histogram of the distribution of maximum peak periods over all iterations. Figures \ref{Kep20_PPev_dist} and \ref{Kep20_Lev_dist} show example histograms for simulations with \emph{$\tau$} = 22.9 days and \emph{$\tau$} = 274.0 days, respectively, both with \emph{$\omega$} = 0.55. All five values of \emph{$\omega$} produce similar final maximum peak distributions, as shown in tables \ref{Kep20_PPev_rates} and \ref{Kep20_Lev_rates}. \\

Tables \ref{Kep20_PPev_rates} and \ref{Kep20_Lev_rates} list occurrence rates for simulations with \emph{$\tau$} = 22.9 days and \emph{$\tau$} = 274.0 days, respectively. We calculated occurrence rates for maximum peaks located within 5$\%$ of \emph{$P_{\rm{rot}}$}, its integer multiples ($P_{\rm{rot}}\times$2, ..., $P_{\rm{rot}}\times$10), its rotational harmonics (\emph{$P_{\rm{rot}}$}/2, ..., \emph{$P_{\rm{rot}}$}/10), the cadence peak (37.3 days, Figure \ref{K20_splt}), and the POI (34.94 days, Figure \ref{K20_splt}). \\
  
\begin{figure*}[]
\begin{center}
\includegraphics[scale=0.35]{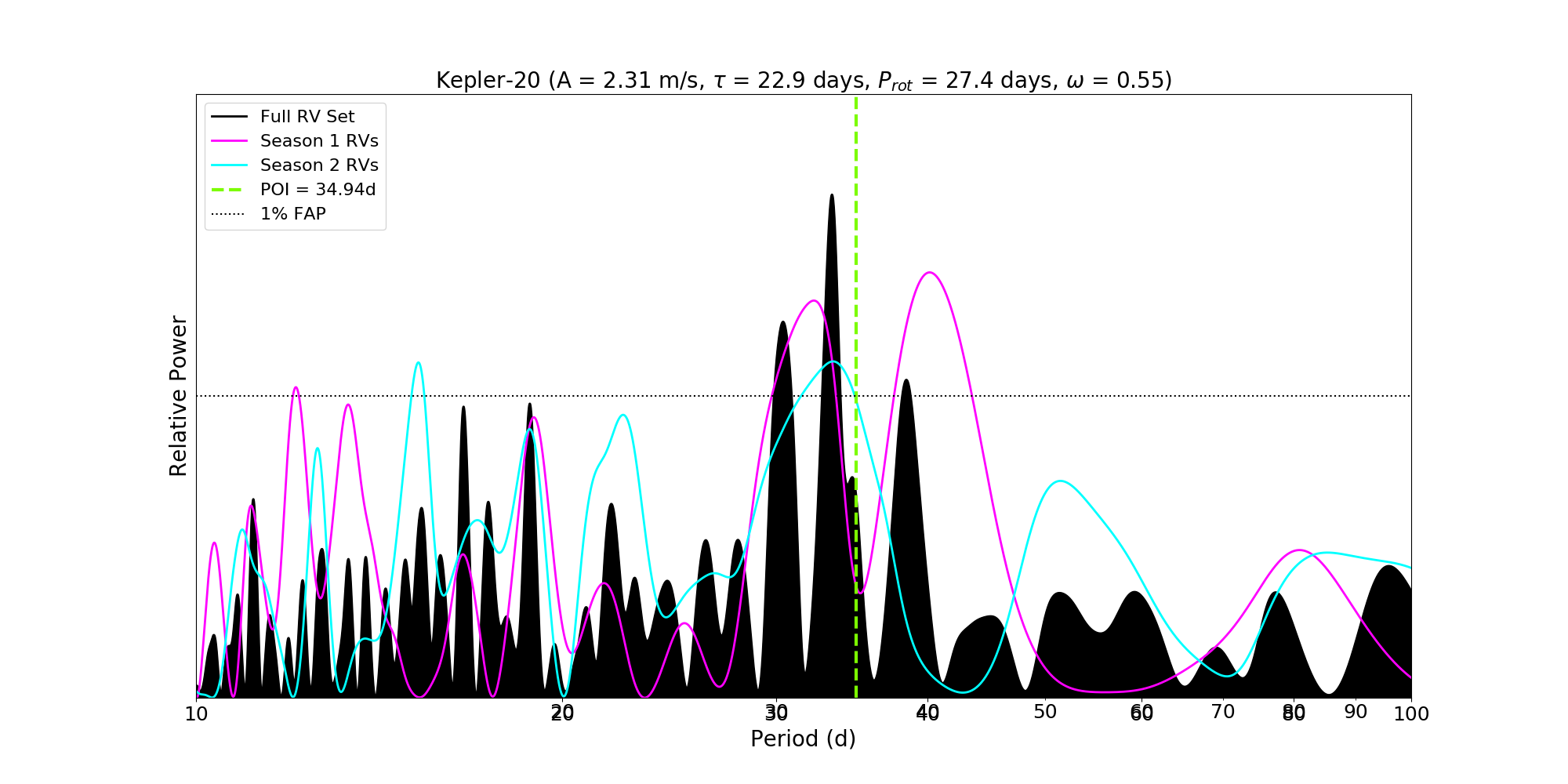}
\end{center}
\caption{An example of simulated Kepler-20 magnetic activity RVs with evolving activity features (\emph{$\tau$} = 22.9 days) and \emph{$\omega$} = 0.55, producing a long-lived peak in the periodogram at the (\emph{POI}), 34.94 days. The power corresponding to a 1\% false alarm probability (FAP) rate is indicated by the dashed black line, and the GLS-periodogram of simulated Kepler-20 RVs is shown in solid black. The magenta and cyan curves show GLS periodograms of just first-season and second-season observations, respectively. We use the same method to calculate periodograms of individual seasons as used in Figure \ref{K20_splt} \citep{Dumusque12}. Simulating only magnetic activity, we are able to reproduce a long-lived signal that could be attributed to a non-transiting planet. Of the maximum peaks occurring at 34.94 days, 16.5\% - 18.7\% were long-lived.}\label{K20_tc}
\end{figure*}

 In iterations with a maximum peak at the POI, we checked whether the maximum peak was long-lived, surviving both seasons of observation. We used the method described in the final paragraph of Section \ref{maxpeak} to calculate a rate of time-coherence at the POI, listed in tables \ref{Kep20_PPev_rates} and \ref{Kep20_Lev_rates}. Figure \ref{K20_tc} shows an example periodogram in which simulated Kepler-20 activity RVs, with \emph{$\omega$} = 0.55, produce a long-lived peak at the POI. \\


\begin{deluxetable*}{cccccc}
\tabletypesize{\scriptsize}
\tablewidth{0pt}
\tablecolumns{7}
\tablecaption{Occurrence rates (\%) for select maximum peak values in simulated RV periodograms of Kepler-20, with evolving activity features (\emph{$\tau$} = 22.9 days). First rates listed are with respect to only iterations having statistically significant peaks, and second rates listed are with respect to all 100,000 iterations.  
\label{Kep20_PPev_rates}}

\tablehead{\colhead{Peak value}  & \colhead{\emph{$\omega$} = 0.45} & \colhead{\emph{$\omega$} = 0.475}  & \colhead{\emph{$\omega$} = 0.5}  & \colhead{\emph{$\omega$} = 0.525}  &  \colhead{\emph{$\omega$} = 0.55}}

\startdata 
POI = 34.94d & 4.3 / 0.7  & 4.7 / 0.8 & 4.4 / 0.8  & 4.8 / 0.9  & 4.8 / 0.9  \\
POI time-coherence & 17.6 & 17.2 & 16.5 & 18.7 & 17.3 \\
\\
\emph{$P_{\rm{rot}}$} = 27.4d  & 6.9 / 1.2  & 6.4 / 1.1 & 6.7 / 1.2   & 6.7 / 1.2  & 6.5 / 1.2  \\
\\
\emph{$P_{\rm{rot}}$} / 2 = 13.7d  & 4.6 / 0.8 & 4.2 / 0.7 & 3.8 / 0.7  & 3.3 / 0.6  & 3.2 / 0.6  \\
\emph{$P_{\rm{rot}}$} / 3 = 9.1d & 1.3 / 0.2  & 1.1 / 0.2  & 1.0 / 0.2  & 0.9 / 0.2  & 0.7 / 0.2  \\
\emph{$P_{\rm{rot}}$} / 4 = 6.9d & 0.5 / 0.1  & 0.5 / 0.1  & 0.4 / 0.1  & 0.3 / 0.1  & 0.4 / 0.1  \\
\emph{$P_{\rm{rot}}$} / 5 = 5.5d & 0.2 / 0.0  & 0.2 / 0.0 & 0.1 / 0.0  & 0.2 / 0.0  & 0.1 / 0.0  \\
\emph{$P_{\rm{rot}}$} / 6 = 4.6d & 0.2 / 0.0  & 0.1 / 0.0  & 0.1 / 0.0  & 0.1 / 0.0  & 0.1 / 0.0  \\
\emph{$P_{\rm{rot}}$} / 7 = 3.9d & 0.1 / 0.0  & 0.1 / 0.0  & 0.1 / 0.0  & 0.1 / 0.0  & 0.1 / 0.0  \\
\emph{$P_{\rm{rot}}$} / 8 = 3.4d & 0.2 / 0.0  & 0.1 / 0.0  & 0.1 / 0.0 & 0.1 / 0.0  & 0.1 / 0.0  \\
\emph{$P_{\rm{rot}}$} / 9 = 3.0d  & 0.1 / 0.0   & 0.2 / 0.0  & 0.1 / 0.0  & 0.1 / 0.0  & 0.1 / 0.0  \\
\emph{$P_{\rm{rot}}$} / 10 = 2.7d & 0.1 / 0.0  & 0.1 / 0.0  & 0.1 / 0.0  & 0.1 / 0.0  & 0.1 / 0.0  \\
\\
\emph{$P_{\rm{rot}}$} x 2 = 54.8d  & 2.3 / 0.4  & 2.0 / 0.3  & 2.1 / 0.4  & 1.5 / 0.5  & 2.4 / 0.5  \\
\emph{$P_{\rm{rot}}$} x 3 = 82.2d  & 1.7 / 0.3  & 1.7 / 0.3   & 1.6 / 0.3  & 1.9 / 0.3  & 1.8 / 0.3  \\
\emph{$P_{\rm{rot}}$} x 4 = 109.6d  & 1.9 / 0.3  & 1.9 / 0.3   & 2.1 / 0.4  & 2.3 / 0.4  & 2.3 / 0.4  \\
\emph{$P_{\rm{rot}}$} x 5 = 137.0d  & 1.6 / 0.3  & 1.7 / 0.3   & 1.8 / 0.3  & 2.2 / 0.4  & 2.0 / 0.4  \\
\emph{$P_{\rm{rot}}$} x 6 = 164.4d  & 1.5 / 0.3  & 1.4 / 0.3  & 1.7 / 0.3  & 2.0 / 0.4  & 2.0 / 0.4  \\
\emph{$P_{\rm{rot}}$} x 7 = 191.8d  & 1.4 / 0.2  & 1.6 / 0.3  & 1.7 / 0.3 & 1.9 / 0.4  & 1.9 / 0.4  \\
\emph{$P_{\rm{rot}}$} x 8 = 219.2d  & 1.5 / 0.3  & 1.6 / 0.3   & 1.9 / 0.4  & 1.9 / 0.4  & 2.2 / 0.4  \\
\emph{$P_{\rm{rot}}$} x 9 = 246.6d  & 1.3 / 0.2  & 1.2 / 0.2   & 1.3 / 0.2  & 1.6 / 0.3  & 1.5 / 0.3  \\
\emph{$P_{\rm{rot}}$} x 10 = 274.0d  & 0.8 / 0.1  & 0.9 / 0.2   & 1.0 / 0.2  & 1.0 / 0.2  & 1.0 / 0.2  \\
\\
cadence peak = 37.3d & 3.1 / 0.6  & 3.1 / 0.5  & 3.0 / 0.5  & 3.2 / 0.6  & 3.3 / 0.6 \\
\enddata
\end{deluxetable*}

\begin{deluxetable*}{cccccc}
\tabletypesize{\scriptsize}
\tablewidth{0pt}
\tablecolumns{7}
\tablecaption{Occurrence rates (\%) as reported in table \ref{Kep20_PPev_rates}, but with stable activity features (\emph{$\tau$} = 274.0 days). \label{Kep20_Lev_rates}}

\tablehead{\colhead{Peak value}  & \colhead{\emph{$\omega$} = 0.45} & \colhead{\emph{$\omega$} = 0.475}  & \colhead{\emph{$\omega$} = 0.5}  & \colhead{\emph{$\omega$} = 0.525}  &  \colhead{\emph{$\omega$} = 0.55}}

\startdata 
POI = 34.94d & 1.1 / 0.3  & 1.3 / 0.4 & 1.5 / 0.4  & 1.7 / 0.5  & 1.7 / 0.5  \\
POI time-coherence & 5.5 & 9.0 & 7.5 & 6.7 & 8.0 \\
\\
\emph{$P_{\rm{rot}}$} = 27.4d & 66.2 / 18.3  & 68.5 / 19.4  & 70.7 / 20.2 & 72.9 / 20.9  & 74.1 / 21.2  \\
\\
\emph{$P_{\rm{rot}}$} / 2 = 13.7d & 16.8 / 4.7 & 14.4 / 4.1 & 12.4 / 3.5  & 10.5 / 3.0  & 9.0 / 2.6  \\
\emph{$P_{\rm{rot}}$} / 3 = 9.1d & 1.4 / 0.4  & 0.9 / 0.3    & 0.7 / 0.2  & 0.5 / 0.1 & 0.4 / 0.1  \\
\emph{$P_{\rm{rot}}$} / 4 = 6.9d & 0.1 / 0.0  & 0.1 / 0.0  & 0.1 / 0.0   & 0.1 / 0.0  & 0.1 / 0.0  \\
\emph{$P_{\rm{rot}}$} / 5 = 5.5d & 0.1 / 0.0  & 0.1 / 0.0  & 0.1 / 0.0  & 0.1 / 0.0  & 0.1 / 0.0  \\
\emph{$P_{\rm{rot}}$} / 6 = 4.6d & 0.1 / 0.0  & 0.1 / 0.0  & 0.1 / 0.0  & 0.1 / 0.0  & 0.1 / 0.0  \\
\emph{$P_{\rm{rot}}$} / 7 = 3.9d & 0.1 / 0.0  & 0.1 / 0.0  & 0.0 / 0.0  & 0.1 / 0.0  & 0.1 / 0.0  \\
\emph{$P_{\rm{rot}}$} / 8 = 3.4d & 0.1 / 0.0  & 0.1 / 0.0  & 0.1 / 0.0  & 0.1 / 0.0  & 0.1 / 0.0  \\
\emph{$P_{\rm{rot}}$} / 9 = 3.0d  & 0.1 / 0.0 & 0.1 / 0.0  & 0.1 / 0.0  & 0.1 / 0.0  & 0.1 / 0.0  \\
\emph{$P_{\rm{rot}}$} / 10 = 2.7d & 0.0 / 0.0  & 0.1 / 0.0 & 0.1 / 0.0  & 0.1 / 0.0  & 0.1 / 0.0  \\
\\
\emph{$P_{\rm{rot}}$} x 2 = 54.8d   & 0.2 / 0.1  & 0.1 / 0.0 & 0.2 / 0.1  & 0.1 / 0.0  & 0.1 / 0.0  \\
\emph{$P_{\rm{rot}}$} x 3 = 82.2d   & 0.1 / 0.0  & 0.1 / 0.0 & 0.1 / 0.0  & 0.1 / 0.0  & 0.1 / 0.0  \\
\emph{$P_{\rm{rot}}$} x 4 = 109.6d  & 0.4 / 0.1  & 0.4 / 0.1 & 0.3 / 0.1  & 0.3 / 0.1  & 0.3 / 0.1  \\
\emph{$P_{\rm{rot}}$} x 5 = 137.0d  & 0.1 / 0.0  & 0.1 / 0.0 & 0.0 / 0.0  & 0.1 / 0.0  & 0.1 / 0.0  \\
\emph{$P_{\rm{rot}}$} x 6 = 164.4d  & 0.0 / 0.0  & 0.0 / 0.0 & 0.0 / 0.0  & 0.0 / 0.0  & 0.0 / 0.0  \\
\emph{$P_{\rm{rot}}$} x 7 = 191.8d  & 0.0 / 0.0  & 0.0 / 0.0 & 0.0 / 0.0  & 0.0 / 0.0  & 0.0 / 0.0  \\
\emph{$P_{\rm{rot}}$} x 8 = 219.2d  & 0.0 / 0.0  & 0.0 / 0.0 & 0.0 / 0.0  & 0.0 / 0.0  & 0.0 / 0.0  \\
\emph{$P_{\rm{rot}}$} x 9 = 246.6d  & 0.1 / 0.0  & 0.1 / 0.0 & 0.1 / 0.0  & 0.1 / 0.0  & 0.1 / 0.0  \\
\emph{$P_{\rm{rot}}$} x 10 = 274.0d & 0.1 / 0.0  & 0.1 / 0.0 & 0.1 / 0.0  & 0.1 / 0.0  & 0.1 / 0.0  \\
\\
cadence peak = 37.3d & 0.6 / 0.2  & 0.7 / 0.2  & 0.9 / 0.3  & 0.9 / 0.2  & 0.9 / 0.2 \\
\enddata
\end{deluxetable*}


\section{Results}\label{results}

Our analyses yield results that provide insight into the interpretation of RV periodogram results with respect to magnetic activity, given the limited, non-uniform sampling typical of current RV observing strategies. \\

\subsection{Unexpected Maximum Peaks in RV Periodograms}\label{peak_wrong}

Figures \ref{K2-131_PPev_dist} and \ref{Kep20_PPev_dist} show final distributions for simulated RVs of K2-131 and Kepler-20 with evolving magnetic active regions, and therefore stellar surfaces that change over the timescale of observations ($\tau \approx P_{\rm{rot}}$). In these cases, a large fraction of simulated active region signals fail to produce a maximum peak at \emph{$P_{\rm{rot}}$} or a related period. Some simulations fail to produce significant peaks at all. This is rarely the case for K2-131, with only 0.1\% - 0.2\% of iterations lacking significant peaks, but in the case of Kepler-20, a whopping 81.0\% - 83.1\% of simulated signals fail to produce significant peaks. The range in reported rates is a result of varying \emph{$\omega$} values, while for simplicity, figures are only shown for a single average active region distribution (\emph{$\omega$} = 0.55). 
A large fraction of the statistically significant maximum peaks that do occur in the RV periodogram are located at periods unrelated to \emph{$P_{\rm{rot}}$}: 80.6\% - 81.0\% of maximum peaks for K2-131 and 71.5\% - 73.0\% of maximum peaks in the case of Kepler-20 (Tables \ref{K2-131_PPev_rates} and \ref{Kep20_PPev_rates}). If analyses of real RV data disregard stellar active regions or use inadequate models, these spurious periodic signals could interact with Keplerian signals of known exoplanets and lead to inaccurate RV mass measurements with underestimated errors.\\

Figures \ref{K2-131_Lev_dist} and \ref{Kep20_Lev_dist} show final distributions for simulated RVs of K2-131 and Kepler-20 with magnetic active regions, and therefore stellar surfaces, that remain unchanged over the timescale of observations (activity evolution timescale increased to $\tau >> P_{\rm{rot}}$). In these cases, where strong rotation signals would typically be expected, a large fraction of simulated activity signals fail to produce a maximum peak at \emph{$P_{\rm{rot}}$} or a related period. While essentially all iterations of simulated K2-131 signals produced significant peaks, simulated Kepler-20 RVs still fail to produce significant peaks at any period in 71.6\% - 72.4\% of iterations. Again, a considerable fraction of the significant maximum peaks that do occur in the RV periodogram are located at periods unrelated to \emph{$P_{\rm{rot}}$}: 34.3\% - 38.2\% of peaks in the case of K2-131 and 14.0\% - 15.0\% of peaks for Kepler-20 (Tables \ref{K2-131_Lev_rates} and \ref{Kep20_Lev_rates}). These results demonstrate that spurious peaks are inherent to RVs of many activity distributions, even when magnetic surface features are unchanged over the timescale of observations. Therefore, even with high-cadence observations of exoplanets orbiting relatively inactive stars, spurious periodic signals could still lead to aforementioned errors in RV mass determinations.\\

\subsection{Observational Cadence and the Stellar Rotation Signal}\label{cad_peak}

All ten of our final distributions show maximum peaks in the simulated RV periodograms favoring a period located between \emph{$P_{\rm{rot}}$} and the cadence peak. In cases where $\tau \approx P_{\rm{rot}}$ (Figures \ref{K2-131_PPev_dist} and \ref{Kep20_PPev_dist}), this feature in the histogram could be attributed to random variations in the overall distributions. However, cases where $\tau >> P_{\rm{rot}}$ (Figures \ref{K2-131_Lev_dist} and \ref{Kep20_Lev_dist}) show clear features in the histogram between \emph{$P_{\rm{rot}}$} and the cadence peak. This suggests that maximum signals in RV periodograms have a tendency to occur at a period related to the limited sampling of the signal at $P_{\rm{rot}}$. In the case of Kepler-20, this favored period occurs at the POI, 34.94 days. \\

\subsection{Occurrence at the POI for Kepler-20 and K2-131}\label{imposters}
For simulations of K2-131 using a real estimate of the activity evolution timescale ($\tau \approx P_{\rm{rot}}$), 1.0\% - 1.7\% of significant maximum peaks in the RV periodogram occurred at the POI, 3.0 days (Table \ref{K2-131_PPev_rates}). Considering the close proximity of the POI to the second rotational harmonic (\emph{$P_{\rm{rot}}$}/3 = 3.2 days), these occurrence rates are strikingly low relative to occurrence rates at other period values. However, as mentioned in section \ref{K2_analysis}, simulations including the lowest value of \emph{$P_{\rm{rot}}$} within its reported uncertainty lead to an overlap between the POI and \emph{$P_{\rm{rot}}$}/3, within the allowed 5\% window. In this case,  occurrence rates at the POI increase to 2.5\% - 3.6\%. Due to the large range of potential maximum peak values in a given distribution, maximum peaks aren't particularly likely to occur at any given period value. Therefore, while occurrence rates at the POI seem low, they appear more significant when compared to the highest occurrence rates at \emph{$P_{\rm{rot}}$}, 6.6\% - 7.7\% (Table \ref{K2-131_PPev_rates}).\\

Simulations of Kepler-20 using a real estimate of the activity evolution timescale ($\tau \approx P_{\rm{rot}}$) produce a maximum peak in the RV periodogram at the POI in 4.3\% - 4.8\% of iterations. These rates are relatively high when compared to the highest occurrence rates at \emph{$P_{\rm{rot}}$}, 6.4\% - 6.9\% (Table \ref{Kep20_PPev_rates}). Occurrence rates at the POI in simulations for Kepler-20 also seem more significant when compared with occurrence rates at the POI in similar simulations for K2-131. Since the POI for K2-131 is relatively close (0.2 days away) to its nearest \emph{$P_{\rm{rot}}$} relative (\emph{$P_{\rm{rot}}$}/3), and the POI for Kepler-20 is more than seven days away from \emph{$P_{\rm{rot}}$}, we would expect to see greater occurrence rates at the POI in the case of K2-131. However, simulations of K2-131 and K2-131b produce much lower relative occurrence rates at the POI, 3.0 days. Our results therefore prove contrary to the assumption that maximum peaks from magnetic activity in RV periodograms usually occur at periods related to \emph{$P_{\rm{rot}}$}. \\

Simulated Kepler-20 RVs further defy assumptions about magnetic activity signals with 16.5\% - 18.7\% of maximum peaks occurring at 34.94 days being long-lived, remaining the maximum peak in periodograms of both seasons of observation. These long-lived periodic signals that occur many days from a star's estimated \emph{$P_{\rm{rot}}$} could be misinterpreted as exoplanet signals, particularly when analyses disregard models for stellar activity. While our results alone cannot rule out or confirm the existence of non-transiting planets around any target for certain, simulations of Kepler-20 demonstrate how spurious signals in RV periodograms from magnetic active regions could appear planetary in nature. \\

\section{Discussion}\label{discussion}

Our ability to detect RV signals of low-mass and long-period planets is currently limited by magnetic activity effects on stellar surfaces, which ubiquitously appear as m/s level variations in even the least active stars \citep[e.g.][]{I&F10}. The key to breaking this magnetic activity barrier is understanding how activity effects appear in observations and finding optimal ways to characterize and model them. Periodogram-based approaches are highly common attempts to distinguish between signals from exoplanets and magnetic active regions. However, the effectiveness of characterizing evolving, quasi-periodic signals with perfectly periodic sine curves is untested. The simulations we present in this paper are a first test of the reliability of common assumptions about magnetic activity signal behavior in RV periodograms. Here we highlight the implications of our results for past and future exoplanet detections and characterizations.\\

The assumption that magnetic activity signals will peak at a period related to \emph{$P_{\rm{rot}}$} in the RV periodogram could lead to inaccurate mass measurements and missed exoplanet signals. Our results in Section \ref{peak_wrong} reveal that RV signals from magnetic active regions often peak at periods unrelated to \emph{$P_{\rm{rot}}$} in the GLS-periodogram, even in the case of high-cadence observations of star's with highly stable, unchanging magnetic regions. These spurious periodic peaks are unlikely to be attributed to magnetic activity when a prior estimate of \emph{$P_{\rm{rot}}$} is known, and they can disguise RV semi-amplitudes of real exoplanet signals. Both targeted follow-up observations to determine masses of known, transiting exoplanets and blind observations to detect new exoplanets are susceptible to this effect. RV fits without a physically motivated model for stellar activity rely heavily on periodogram analyses and risk inaccurately measuring masses of known transiting exoplanets, or missing/misidentifying signals of unknown companions completely.\\

Magnetic activity signals produce long-lived peaks that could be misidentified as non-transiting companions. Results discussed in Section \ref{imposters} reveal multiple examples of a purely quasi-periodic simulated magnetic activity cycle producing the same spurious maximum peak in the RV periodogram over multiple seasons of observation. Our results in Section \ref{cad_peak} suggest that these long-lived spurious signals may be related to the limited sampling of the rotation signal, and therefore tend to occur at a period between \emph{$P_{\rm{rot}}$} and the strongest periodic signal inherent to observational sampling, the cadence peak. Given the common assumption that magnetic activity cycles will not produce long-lived significant peaks at periods unrelated to \emph{$P_{\rm{rot}}$}, these long-lived spurious peaks from magnetic activity could be mistakenly attributed to an exoplanet. Fits excluding a model for magnetic activity are particularly vulnerable to misidentifying these spurious periodic signals.\\

In order to model and fit stellar activity signals, we need to utilize methods that provide reliable prior estimates of associated parameters, particularly \emph{$\tau$} and \emph{$P_{\rm{rot}}$}. ACF analyses and GP regression fits applied to Kepler LCs have provided \emph{$\tau$} and \emph{$P_{\rm{rot}}$} estimates leading to successful RV mass measurements \citep[e.g.][]{haywood14,LMH}. Ideally, photometric LCs used to inform magnetic activity models should be observed near the same time frame as RVs, in order to avoid comparing data sets taken at different phases of a star's activity cycle or with dramatically different activity feature distributions. Spectroscopic activity indicators and chromospheric RVs can also constrain magnetic activity fits, and are inherently simultaneous with RV observations.\\

The current approach of RV surveys is to use stellar activity information retroactively, correcting effects from magnetic active regions in RV measurements after data have already been collected. However, a more efficient way to deal with activity effects could be to schedule RV observations of specific systems in ways that optimize the sampling of both the magnetic activity and exoplanet signals, so both can be more easily extracted from the data. Early attempts to do this have proven successful \citep{LMH, barros17, santerne18}. As more precise and stable RV instruments become available, collaboration between those instruments will be necessary to achieve better coverage of stellar activity cycles. Refined algorithms can then produce optimized observing strategies for individual targets. These steps will be key to accurately measuring exoplanet masses in RV data.\\

Our results in Section \ref{results} reveal that magnetic activity cannot be ignored in RV exoplanet fits. A fit accounting for stellar activity should be considered for all potential exoplanet signals, either with GP regression or another well-motivated model. This is true even for cases where stellar activity signals are not obviously observed with strong peaks in the periodogram or correlations to known activity indicators. The simulations detailed in this paper can become a standard tool for determining what signals to expect from magnetic activity in the RV periodogram, comparing those signals with real data sets, and preventing assumptions about peak locations from leading to inaccurate mass measurements and false exoplanet detections.

\section*{Acknowledgements}
We would like to thank Annelies Mortier and Andrew Collier Cameron for sharing their radial velocity and stellar activity expertise in meetings. Some of this work has been carried out in the frame of the National Centre for Competence in Research `PlanetS' supported by the Swiss National Science Foundation (SNSF). This material is based upon work supported by the National Aeronautics and Space Administration under grants No. NNX15AC90G and NNX17AB59G issued through the Exoplanets Research Program. This work was also performed under contract with the California Institute of Technology (Caltech)/Jet Propulsion Laboratory (JPL) funded by NASA through the Sagan Fellowship Program executed by the NASA Exoplanet Science Institute (R.D.H.).

\bibliography{ref.bib}


\clearpage



\clearpage






\end{document}